\documentclass[preprint]{aastex}
\usepackage{mkfig}
\begin{document}

\title{HST Observations of HD 166620 and $\tau$~Ceti:  First UV
  Spectra of a Magnetic Grand Minimum Star and the Extent of
  $\tau$~Ceti's Astrosphere}
%\title{HST Observations of HD 166620 and $\tau$~Ceti:  First UV
%  Spectra of a Maunder Minimum Star and the Origin of
%  $\tau$~Ceti's IR-Bright Debris Disk}
%\title{HST Observations of HD 166620 and $\tau$~Ceti:  First UV
%  Spectra of a Maunder Minimum Star and Inference that $\tau$~Ceti's
%  Debris Disk Extends Into the ISM}
%\title{First UV Spectroscopic Observations of a Star in a Maunder Minimum
%  State}\altaffilmark{1}

\author{Brian E. Wood\altaffilmark{2},
  Hans-Reinhard M\"{u}ller\altaffilmark{3}, Dean Hartshorn\altaffilmark{4},
  Seth Redfield\altaffilmark{4},
  Travis S. Metcalfe\altaffilmark{5}}
\altaffiltext{1}{Based on observations made with the NASA/ESA Hubble
  Space Telescope, obtained at the Space Telescope Science Institute,
  which is operated by the Association of Universities for Research
  in Astronomy, Inc., under NASA contract NAS 5-26555.  These
  observations are associated with program GO-17793.}
%\altaffiltext{1}{DISTRIBUTION A.  Approved for public release:
%  distribution unlimited}
\altaffiltext{2}{Naval Research Laboratory, Space Science Division,
  Washington, DC 20375, USA; brian.e.wood26.civ@us.navy.mil}
\altaffiltext{3}{Department of Physics and Astronomy, Dartmouth College,
  Hanover, NH 03755, USA}
\altaffiltext{4}{Astronomy Department and Van Vleck Observatory,
  Wesleyan University, Middletown, CT 06459-0123, USA}
\altaffiltext{5}{Center for Solar-Stellar Connections, WDRC,
  9020 Brumm Trail, Golden, CO 80403, USA}

%\linenumbers

\begin{abstract}

     We present new {\em Hubble Space Telescope} (HST) UV spectra of
the K2~V star HD~166620, the first star clearly recognized to be in
a ``magnetic grand minimum'' state analogous to the Sun's ``Maunder
Minimum'' in the late 1600's.  The stellar H~I Lyman-$\alpha$
surface fluxes are extremely low, about a factor of two below
fluxes observed during solar minimum, and also significantly
lower than those of $\tau$~Ceti (G8~V) and HD~191408 (K2.5~V), two
stars more similar to HD~166620 in spectral type and age ($\sim 10$~Gyr)
than the Sun.  The $\tau$~Ceti data that are compared with HD~166620
include both old archival data and a new HST observation as well.
The Lyman-$\alpha$ data are used to confirm a nondetection of
astrospheric Lyman-$\alpha$ absorption for this star, suggesting a
very weak wind with $\dot{M}<0.1$~$\dot{M}_{\odot}$.
The very compact astrosphere inferred for $\tau$~Ceti indicates
that the star's debris disk is at least partly exposed to the ISM,
and we discuss possible consequences.

\end{abstract}


\section{Introduction}

     Our Sun has a well known 11-year activity cycle during which
it alternates between periods of high activity, with lots of
sunspots, active regions, and accompanying chromospheric and coronal
activity; and periods of low activity, with no or few spots and
active regions.  This behavior is governed by the Sun's internal
magnetic dynamo, rooted in the Sun's outer convection zone, where
convective plasma motions influenced by differential rotation are
generating magnetic fields \citep{pc20}.  One avenue for
exploring the nature of the activity cycle involves observing the
cyclic behavior of other stars, which can be particularly important
for inferring what the solar cycle may have been like in the past
and what it might be like in the future.

     Since stellar dynamos are driven in part by stellar rotation,
the evolution of activity cycles over time will be tied to the
evolution of stellar rotation.  Stars begin their lives rotating
rapidly, but this rotation slows with time due to a magnetic braking
process whereby stellar winds drag against the large-scale field
of the star, allowing stars to gradually shed angular momentum.
The {\em Kepler} mission radically increased the number of stars
with measured rotation periods, resulting in the surprising
discovery that magnetic braking appears to slow or even shut down
for old stars \citep{jlvs16,jlvs19,ojh21,tjd22}.
There is evidence that this weakened magnetic
braking appears to be accompanied by an increasing activity cycle
period, until a point where the activity cycle disappears entirely
\citep{tsm17,tsm22}.

     The Sun itself appears to be in this transitional state,
lying above the relation between cycle period and rotation period
established by younger stars with rotation rates that are
still slowing.  A glimpse into the Sun's future as a non-cycling,
flat-activity star is possibly provided by the ``magnetic grand minimum''
(MGM) phenomenon.  During an MGM phase, the Sun's activity cycle
shuts down entirely.  By far the most well known MGM is
the ``Maunder Minimum,'' an extended period from roughly 1645--1715
when sunspots virtually disappeared \citep{jae76,igu00},
but other MGM periods have been identified over the last few
millenia, using isotopic abundance proxies for sunspot number
\citep{igu21,igu25}.  It is possible that MGM phases are
a direct symptom of weakened magnetic braking, and an activity cycle
that is in the process of gradually disappearing \citep{tsm25b}.

     A significant number of stars monitored for activity cycles have
been found to have low, flat activity curves.  These stars had
originally been interpreted to possibly be in MGM states
\citep{sb90}, but most are now generally assumed
to be inactive subgiants or very old main sequence stars with
cycles that have simply disappeared entirely \citep{jtw04,tsm22}.
Identifying a truly convincing MGM star requires monitoring for
sufficiently long enough to discern clear cyclic behavior, followed
by a clear cessation of that cyclic behavior for an extended
period of time.  This naturally requires many decades of observation.
The first definitive MGM star appears to be the K2~V star HD~166620,
which was observed to have a $\sim 17$~yr period from 1966--2004,
but which has had low, flat activity for over 20 years since then
\citep{acb22,jkl22}.  Similar to our Sun,
HD~166620 lies above the cycle-period/rotation-period relation
defined by stars still experiencing full magnetic braking
\citep{tsm22}, consistent
with the hypothesis that MGM phases may be a phenomenon for stars
with weakened magnetic braking, and are precursors to a complete
shut down of the activity cycle.

     We here present UV spectra of HD~166620 from the Space Telescope
Imaging Spectrograph (STIS) instrument on board the {\em Hubble
Space Telescope} (HST), in order to provide the first UV measurements
for a star in an MGM state.  These observations can hopefully provide
some indication of what the Sun's UV emission may have been like
during its MGM phases, particularly the famous Maunder Minimum.
However, there is a significant spectral type difference between
the Sun and HD~166620.  Thus, we will mostly be comparing our
HD~166620 observations to spectra of $\tau$~Ceti (G8~V) and
HD~191408 (K2.5~V), which are closer in size, mass, and spectral type
to HD~166620.  Like HD~166620, HD~191408 is a very old, inactive early
K dwarf, where we will be taking advantage of a recently completed
study of the star's H~I Lyman-$\alpha$ line from \citet{ay22}.
As for $\tau$~Ceti, it is an excellent example of an old,
flat-activity star that has not shown a hint of an activity cycle
since monitoring first began in the late 1960s \citep{slb95,acb22}.
Thus, in comparing HD~166620 with $\tau$~Ceti,
we will be comparing an MGM star with a temporarily halted activity
cycle to a similar star that does not seem to have any cycle at all.
For $\tau$~Ceti, we will not only utilize old, archival HST/STIS
spectra, but we have also obtained more recent HST data that
can be presented here for the first time.

\section{HST Observations}

%\begin{deluxetable}{cccccc}
%\tabletypesize{\scriptsize}
%\tablecaption{HST/STIS Observations}
%\tablecolumns{5}
%\tablewidth{0pt}
%\tablehead{
%  \colhead{Star} &\colhead{Start Time}&
%    \colhead{Grating}&\colhead{Wavelengths (\AA)} & \colhead{Exp. Time (s)}}
%\startdata
\begin{table}[t]
\begin{center}
%\caption{HST/STIS Targets}
Table 1: HST/STIS Observations
\small
\begin{tabular}{ccccc} \hline\hline
Star & Start Time & Grating & Wavelengths (\AA) & Exp. Time (s) \\
\hline
HD 166620  & 2025-05-18 08:15:32 & E230H & 2574--2851 & 1543 \\
           & 2025-05-18 09:30:41 & E140M & 1150--1700 & 2629 \\
$\tau$ Cet & 2022-08-19 19:27:18 & E230H & 2574--2851 & 300 \\
           & 2000-08-01 02:59:12 & E140M & 1150--1700 & 13450 \\
           & 2025-06-15 14:12:02 & E140M & 1150--1700 & 2093 \\
HD 191408  & 2018-09-04 12:12:30 & E230H & 2574--2851 & 306 \\
           & 2018-09-04 10:55:44 & E140M & 1150--1700 & 4064 \\
\hline
\end{tabular}
\end{center}
%\scriptsize
%\normalsize
\end{table}
%\enddata
%\end{deluxetable}
     The UV emission lines that we will be focusing on are primarily
the H~I Lyman-$\alpha$ (Ly$\alpha$) line at 1215.7~\AA\ and the
Mg~II h \& k lines at 2803.5~\AA\ and 2796.3~\AA, respectively.
These are the strongest chromospheric lines available in the UV,
and as such are the best diagnostics of chromospheric activity.
Table~1 lists the HST/STIS observations considered here, consisting
of both new and archival data.  These spectra include near-UV (NUV)
spectra of the 2574--2851~\AA\ spectral region with the E230H
grating, providing the highest possible spectral resolution of
the Mg~II lines, and far-UV (FUV) spectra of the 1150--1700~\AA\
region with the moderate resolution E140M grating.

     As shown in Table~1, the E230H and E140M spectra of HD~166620
are recent observations from 2025~May~18.  In the same observing
program, we obtained a new $\tau$~Ceti E140M spectrum on
2025~June~15, though we will also be considering the older E140M and
E230H spectra from 2000~August~1 and 2022~August~19, respectively.
Table~1 also lists E230H and E140M spectra of the K2.5~V star
HD~191408 from 2018~September~4, as the Ly$\alpha$ and
Mg~II lines of this star will also be utilized, for reasons
that will be made clear in the next section.  All spectra were
initially processed using the standard HST/STIS CALSTIS software,
though as we will discuss in detail in Section~4, modifications to
the data reduction were found necessary to properly extract some of
the weaker lines in the E140M spectra.
\begin{table}[t]
\begin{center}
%\caption{HST/STIS Targets}
Table 2: Stellar Properties
\scriptsize
\begin{tabular}{cccccccccccc} \hline\hline
Star & Spect. & Dist. &  Radius   & V$_{rad}$   & P$_{rot}$ & P$_{cyc}$ &
  [Fe/H] & [Mg/H] & [C/H] & Age & Refs \\
     & Type   & (pc)  &(R$_{\odot}$)&(km s$^{-1}$)&(days)   & (yrs)    &
         &        &       &(Gyr) &     \\
\hline
$\tau$ Cet & G8 V & 3.65 &0.793& -16.6 & 46 &flat&-0.52&-0.25&-0.39&
  $9.0\pm 1.0$  & 1, 2, 3, 4 \\
HD 191408  &K2.5 V& 6.01 &0.744&-129.3 & ...&... &-0.47&-0.09&-0.39&
  9--14& 4, 5, 6\\
HD 166620  & K2 V & 11.1 &0.782& -19.3 & 45 & 17 &-0.10& 0.11&-0.19&
  $9.5^{+2.15}_{-1.65}$    & 2, 4, 7, 8\\
%
%\begin{tabular}{|lccccccccccccccc|} \hline\hline
%Gliese \#& HD & Other & Spect. & Dist. &  Radius   & P$_{rot}$ & P$_{cyc}$ &
%  [Fe/H] & [Mg/H] & [C/H] & $V_{ISM}$& $\theta$ & $\log L_{x}$ & Age & Refs \\
%         &    & Name  & Type   & (pc)  &(R$_{\odot}$)&(days)   & (yrs)    &
%         &        &       & (km/s)  &    (deg) & (ergs/s)    &(Gyr) &     \\
%\hline
%GJ 71  &  10700 & $\tau$ Cet & G8 V & 3.65 &0.793& 46 &flat&-0.52&-0.25&-0.39&
%  55.9 & 59.5 & 26.69 & 10  & 1, 2, 3, 4 \\
%GJ 783A& 191408 & HR 7703    &K2.5 V& 6.01 &0.744& ...&... &-0.47&-0.09&-0.39&
% 119.4 & 19.9 & 26.92 &9--14& 4, 5, 6\\
% GJ 706 & 166620 & HR 6806    & K2 V & 11.1 &0.782& 45 & 17 &-0.10& 0.11&-0.19&
%  49.3 & 83.1 & 26.96 & 12.4& 3, 4, 7, 8\\
%
%GJ 764 & 185144 &$\sigma$ Dra& K0 V & 5.76 & 72.7 &112.0 & 27.78 & 4.5 & 2\\
%GJ 19  &   2151 & $\beta$ Hyi&G2 IV & 7.46 & 63.3 &116.1 & 27.45 & 6.3 & 2\\
%GJ 434 & 101501 & 61 UMa     & G8 V & 9.58 & 33.5 & 78.8 & 28.35 & 1.4 & 2\\
%GJ 598 & 141004&$\lambda$ Ser& G0 V & 11.9 & 48.1 & 16.6 & 27.79 & 6.7 & 2\\
%GJ 616 & 146233 & 18 Sco     & G2 V & 14.1 & 38.6 &129.5 & 27.23 & 3.7 & 2\\
\hline
\end{tabular}
\end{center}
\scriptsize
References --- (1) \citet{mk23}; (2) \citet{acb22};
  (3) \citet{ykt11}; (4) \citet{cap04};
  (5) \citet{ay22}; (6) \citet{lg10};
  (7) \citet{tsm25a}; (8) \citet{jkl22}.
%References --- (1) Korolik et al.\ (2023); (2) Baum et al.\ (2022);
%  (3) Di Folco et al.\ (2004); (4) Allende Prieto et al.\ (2004);
%  (5) Youngblood et al.\ (2022); (6) Ghezzi et al.\ (2010);
%  (7) Metcalfe et al.\ (2025a); (8) Luhn et al.\ (2022).
\normalsize
\end{table}
%REFS: 1. Korolik et al. 2023, 2. Baum et al. 2022 (Pcyc),
%  3. Di Folco et al. 2004 (age), 4. Allende Prieto et al. 2004 (abunds),
%  5. Youngblood et al. 2022 (rad), 6. Ghezzi et al. 2010 (age),
%  7. Metcalfe et al. 2025 (rad), 8. Luhn et al. 2022 (Prot)
%NOTES: Log(Lx)---GJ706 is HRI, GJ783A is average of 2 pointed ROSAT/PSPC
%NOTES: Prot --- GJ71 very uncertain due to pole-on

     Table~2 lists basic stellar properties that we will be
assuming.  The listed spectral types, distances, and radial
velocities (V$_{rad}$) are from the SIMBAD database.
Photospheric abundances from \citet{cap04} are
listed for three elements:  [Fe/H], [Mg/H], and [C/H].
Following convention, the abundances are listed as logarithmic
values relative to solar, so a value of zero would
imply a solar abundance.  The activity cycle periods, P$_{cyc}$,
are from \citet{acb22}.  \citet{mk23} provide
a recent reassessment of $\tau$~Ceti's properties,
which is our source for the stellar radius and rotation period
(P$_{rot}$) of this star, though $\tau$~Ceti's pole-on
orientation makes it very difficult to infer an unambiguous
rotation period despite extensive photometric monitoring.
For HD~166620, P$_{rot}$ is from \citet{jkl22}.
The stellar radii of HD~191408 and HD~166620 are from
\citet{ay22} and \citet{tsm25a},
respectively.  The estimated ages of the three stars are
from \citet{ykt11}, \citet{lg10}, and \citet{tsm25a}.  For
$\tau$~Ceti, the quoted stellar age ($9.0\pm 1.0$~Gyr) is from
asteroseismology \citep{ykt11}.  For HD~191408, we follow
\citet{ay22} in using the isochrone-derived $9-14$~Gyr age range
from \citet{lg10}.  For HD~166620, we assume $9.5^{+2.15}_{-1.65}$~Gyr
based on gyrochronology, but with correction for weakened magnetic
braking.  All of our stars are old stars
with ages of about 10~Gyr.  It is worth mentioning that all
three have very low X-ray luminosities, with $\log L_X<27.0$
measured by ROSAT \citep[e.g.,][]{js04}, consistent
with old age.

\section{Interstellar Absorption Analysis}

     Figure~1 shows the H~I Ly$\alpha$ lines of our
three target stars.  For $\tau$~Ceti, both the old spectrum
from 2000 and the new one from 2025 are shown.  Using the
distances and stellar radii listed in Table~2, the observed
fluxes have been converted to surface fluxes.  The Ly$\alpha$
line profiles are complicated by extensive ISM absorption.
The Sun lies within a partly neutral cloud called the
Local Interstellar Cloud (LIC), so even for the nearest stars
the Ly$\alpha$ absorption from the ISM is strong and very
broad, typically centered within $\sim 30$ km~s$^{-1}$ of the
heliocentric rest frame.  The ISM also produces narrow
deuterium (D~I) absorption $-81.6$ km~s$^{-1}$ from the H~I
absorption.  The location of the D~I absorption is explicitly
indicated in Figure~1.

     Finally, there is also geocoronal Ly$\alpha$ emission
present in the data.  This narrow emission is indicated by the
shaded regions in Figure~1, mostly contained in the saturated
core of the ISM H~I absorption, making it easy to remove by
fitting a Gaussian to the emission and then subtracting the
Gaussian.  The geocoronal emission should be centered on the
Earth's projected velocity toward the observed star, which is
known.  Thus, this emission is actually an excellent wavelength
calibrator, and we correct the wavelength scale of our
spectra accordingly \citep[see][]{bew05b}.  Note that for
HD~191408 the geocoronal emission is only faintly visible at
about $-23$ km~s$^{-1}$.  It is very weak because this spectrum
was taken through the narrow
$0.2^{\prime\prime}\times 0.06^{\prime\prime}$ slit instead of the
more standard $0.2^{\prime\prime}\times 0.2^{\prime\prime}$ aperture.
This suppresses the geocoronal emission significantly, and
improves spectral resolution somewhat, but it does mean the
flux calibration will be more uncertain, which could be a
concern for our purposes.

     We are mostly interested in the intrinsic stellar
Ly$\alpha$ fluxes and the intrinsic line profiles, but
inferring these from the spectra in Figure~1 requires correcting
for the ISM absorption.  This is somewhat easier for HD~191408
than for the other two stars due to the large radial
velocity (V$_{rad}=-129.3$ km~s$^{-1}$), which blueshifts
the stellar emission away from the ISM rest frame.  This is
a major reason why we have included HD~191408 in our study.
Not only is this star similar in spectral type to HD~166620
and $\tau$~Ceti, and with a similar old age of about 10~Gyr,
but the left half of the stellar profile is relatively free
of ISM absorption.  The Ly$\alpha$ line for HD~191408 has
already been reconstructed by \citet{ay22},
a reconstruction that should be somewhat more accurate than
the ones we will do for $\tau$~Ceti and HD~166620, with more
obscured line profiles.

\begin{figure}[t]
\plotfiddle{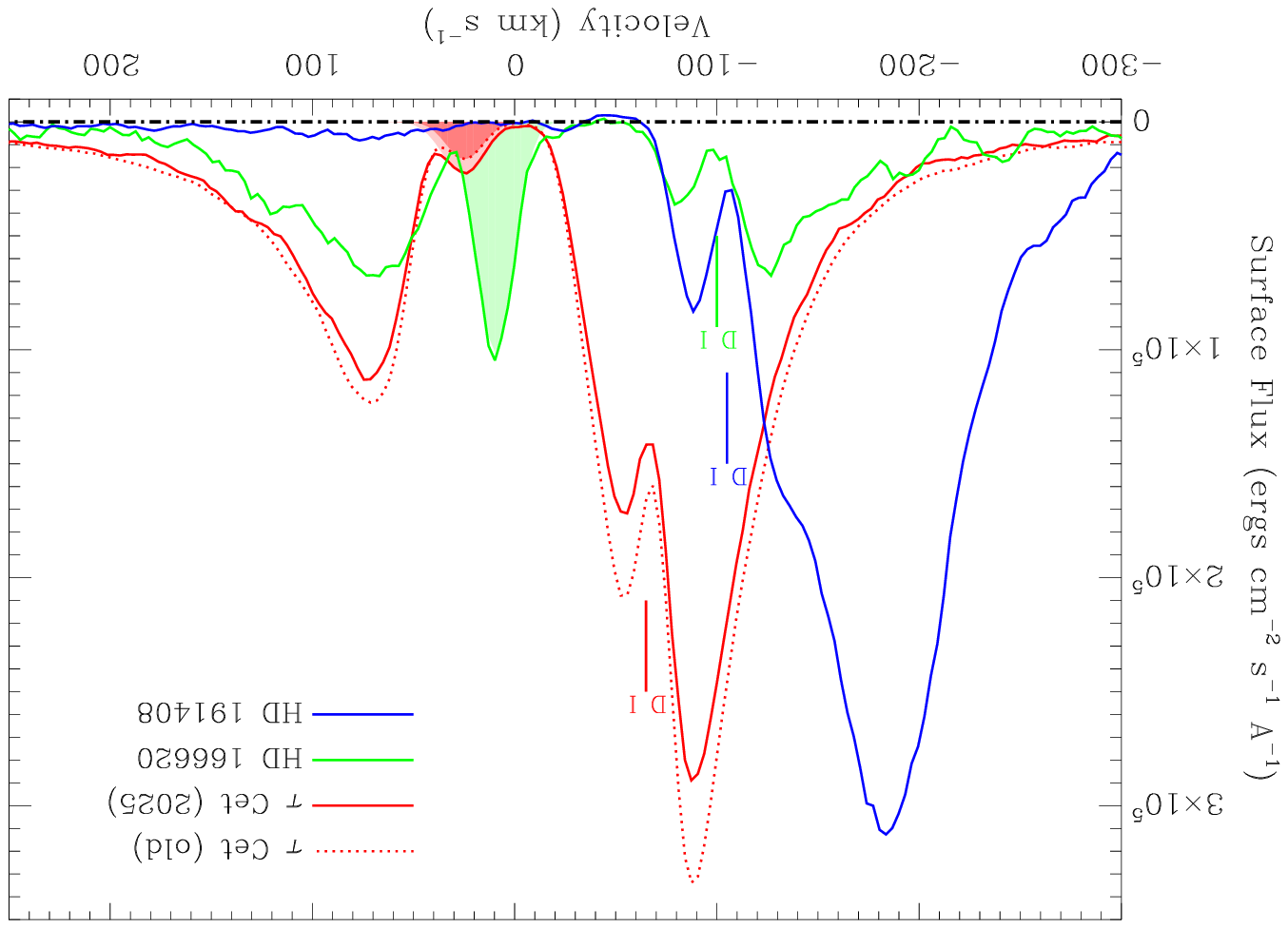}{3.0in}{180}{50}{50}{190}{270}
\caption{Slightly smoothed HST/STIS spectra of the
  H~I Ly$\alpha$ line of three stars, plotted in surface
  flux units on a heliocentric velocity scale.  Two separate
  spectra are shown for $\tau$~Ceti.  Very broad H~I ISM
  absorption absorbs the central part of the line profile,
  with narrow D~I absorption also being present blueward
  of H~I.  Shaded regions indicate narrow geocoronal
  emission within the saturated core of the H~I absorption.}
\end{figure}
     With \citet{bew05b} and \citet{ay22}
already having analyzed the old $\tau$~Ceti and HD~191408
spectra, respectively, we here focus on HD~166620 and the
new $\tau$~Ceti spectrum.  Before tackling
Ly$\alpha$, it is useful to learn more about the ISM
velocity structure by analyzing the much narrower ISM
Mg~II and Fe~II absorption lines in the high resolution
E230H spectra.  \citet{ahn25} have already analyzed
the $\tau$~Ceti lines, allowing us to focus on HD~166620.
Figure~2 shows the ISM Mg~II and Fe~II absorption lines
observed toward HD~166620.

\begin{figure}[t]
\plotfiddle{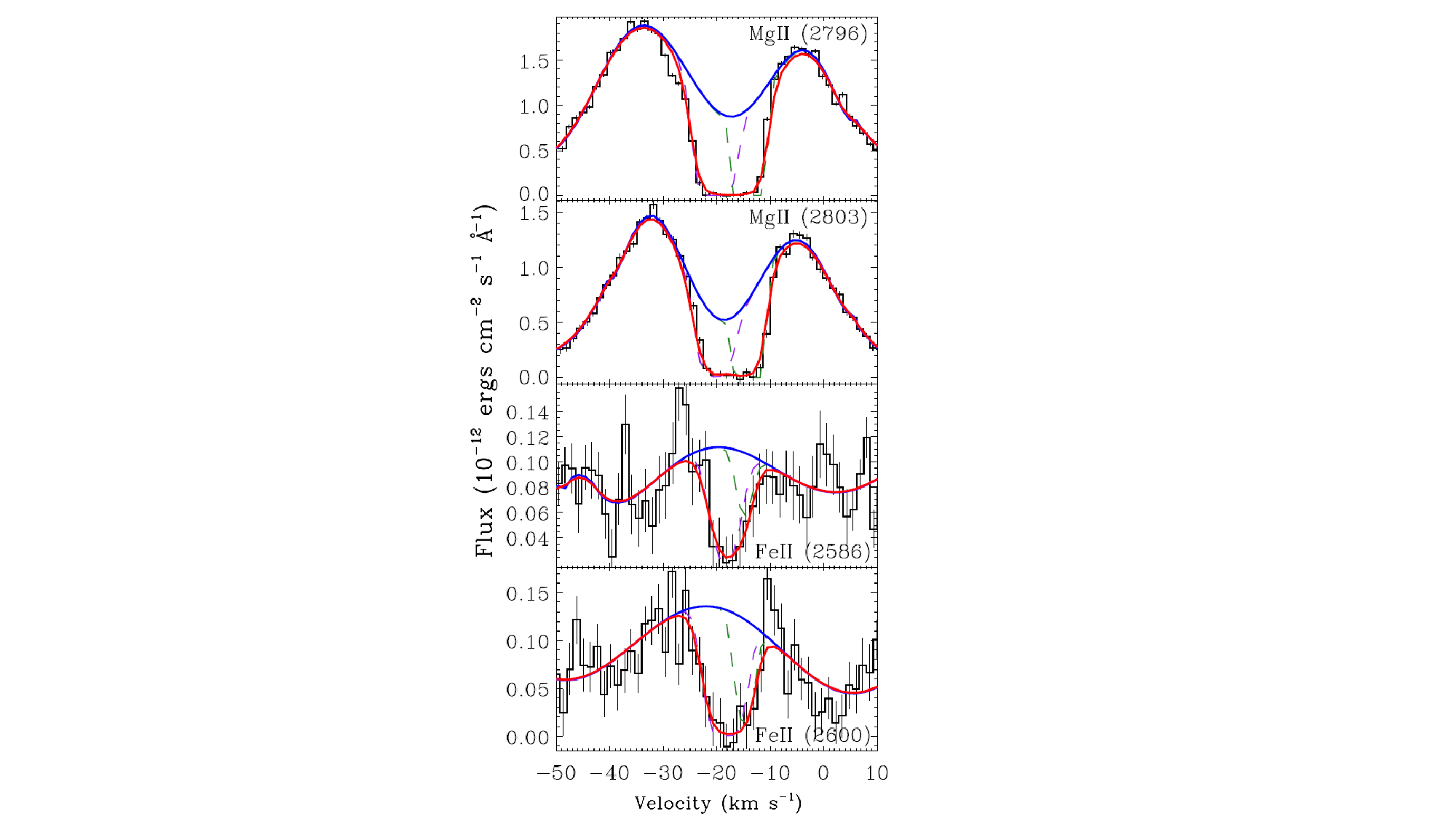}{3.6in}{0}{60}{60}{-280}{-10}
\caption{HST/STIS spectra of ISM Mg~II and Fe~II absorption
  lines observed toward HD~166620, plotted on a heliocentric
  velocity scale.  Blue lines indicate the assumed stellar
  background above the absorption.  The absorption profiles
  are fitted with two components (dashed lines), with
  the red line being the combination of the two after
  instrumental smoothing, which fits the data.}
\end{figure}
     We fit these lines as in past
analyses \citep[e.g.,][]{sr02,sr04,ahn25}.  Each
absorption component is defined by three parameters:  the
central velocity ($v$), Doppler broadening parameter ($b$),
and column density ($N$).  Before comparing a model
absorption profile with the data, it is convolved with
the instrumental line spread function \citep{sm23}.
Although the evidence is subtle, it is clear that there are
at least two ISM velocity components present in the absorption.
Each of the two Mg~II lines is fitted independently, and then
the two lines are fitted simultaneously with parameters forced
to be self-consistent.  In each case, the best two-component
fit is determined by $\chi^2$ minimization, with uncertainties
in the fit parameters estimated using a Monte Carlo technique.
This provides three separate measurements of the ISM
absorption.  The final Mg~II fit parameters reported in Table~3
are a weighted mean of the three independent measurements,
though the fit shown in Figure~2 is the simultaneous fit to
both lines.  The two Fe~II ISM lines are similarly analyzed,
with results also shown in Figure~2 and Table~3.
The $\tau$~Ceti Mg~II and Fe~II line parameters measured
by \citet{ahn25} are listed in Table~3 as well.  Only
one velocity component is observed toward $\tau$~Ceti.
%\begin{deluxetable}{llccccc}
%\tabletypesize{\scriptsize}
%\tablecaption{Absorption Line Fit Parameters\tablenotemark{a}}
%\tablecolumns{7}
%\tablewidth{0pt}
%\tablehead{
%  \colhead{Star}&\colhead{Ion}&\colhead{$\lambda_{rest}$\tablenotemark{b}} &
%    \colhead{ISM} & \colhead{$v$\tablenotemark{c}} & \colhead{$b$} &
%    \colhead{log N} \\ %& \colhead{$\chi^{2}_{\nu}$} \\
%  \colhead{}&\colhead{}&\colhead{(\AA)} & \colhead{Cloud}&\colhead{(km~s$^{-1}$)} &
%    \colhead{(km~s$^{-1}$)} & \colhead{log(cm$^{-2}$)}} % & \colhead{}}
%\startdata
\begin{table}[t]
\begin{center}
%\caption{HST/STIS Targets}
Table 3: ISM Absorption Measurements$^a$
\small
\begin{tabular}{ccccccc} \hline\hline
Star & Ion & $\lambda_{rest}$$^b$ & ISM & $v$$^c$   & $b$      & log N \\
     &     & (\AA)          &Cloud &(km s$^{-1}$)&(km s$^{-1}$)&log(cm$^{-2}$) \\
\hline
HD 166620 & Mg II & 2796.3543, 2803.5315 & Mic &$-20.44\pm 0.70$ &
   $2.64\pm 0.17$ & $12.98\pm 0.13$ \\
          & Mg II & 2796.3543, 2803.5315 & LIC &$-14.19\pm 0.46$ &
   $1.66\pm 0.18$ & $13.92\pm 0.28$ \\
          & Fe II & 2586.6500, 2600.1729 & Mic &$-18.47\pm 0.10$ &
   $3.25\pm 0.51$ & $13.245^{+0.053}_{-0.060}$  \\
          & Fe II & 2586.6500, 2600.1729 & LIC &$-14.81\pm 0.27$ &
   $1.61\pm 0.50$ & $12.744\pm 0.067$  \\
          & H I   &1215.6682, 1215.6736& Mic   &$-19.84\pm 0.97$ &
  $13.03\pm 1.42$ & $18.235\pm 0.021$ \\
          & H I   &1215.6682, 1215.6736& LIC   &($-16.19\pm 0.97$)&
 ($13.03\pm 1.42$)&($17.734\pm 0.021$)\\
$\tau$~Cet (old)$^d$ & Mg II & 2796.3543, 2803.5315 & LIC? & $13.26\pm 0.46$ &
   $3.07\pm 0.62$ & $13.12\pm 0.40$ \\
                 & Fe II & 2586.6500, 2600.1729 & LIC? & $14.29\pm 0.15$ &
   $1.70\pm 0.37$ & $12.92\pm 0.70$ \\
                 & H I   &1215.6682, 1215.6736  & LIC? & $12.34\pm 0.06$ &
   $10.32\pm 0.06$ & $18.006\pm 0.002$ \\
$\tau$~Cet (2025) & H I   &1215.6682, 1215.6736  & LIC? & $12.59\pm 0.14$ &
   $10.20\pm 0.19$ & $18.066\pm 0.005$ \\
\hline
\end{tabular}
\end{center}
\small
Notes --- $^a$Values in parentheses are fixed relative to other component
  (see text).  $^b$Rest wavelengths of measured lines, in vacuum.
  $^c$Central velocity in a heliocentric rest frame.  $^d$From
  \citet{bew05b} and \citet{ahn25}.
%NOTE: 1.1 km/s added to old TauCet(old) H vel. for HI(geo) correction
%NOTE: 0.1 km/s subtracted from TauCet(2025) H vel. for HI(geo) correction
\normalsize
\end{table}
%\enddata
%\tablenotetext{a}{Values in parentheses are fixed relative to other component
%  (see text).}
%\tablenotetext{b}{Rest wavelengths of measured lines, in vacuum.}
%\tablenotetext{c}{Central velocity in a heliocentric rest frame.}
%\end{deluxetable}

\begin{figure}[t]
\plotfiddle{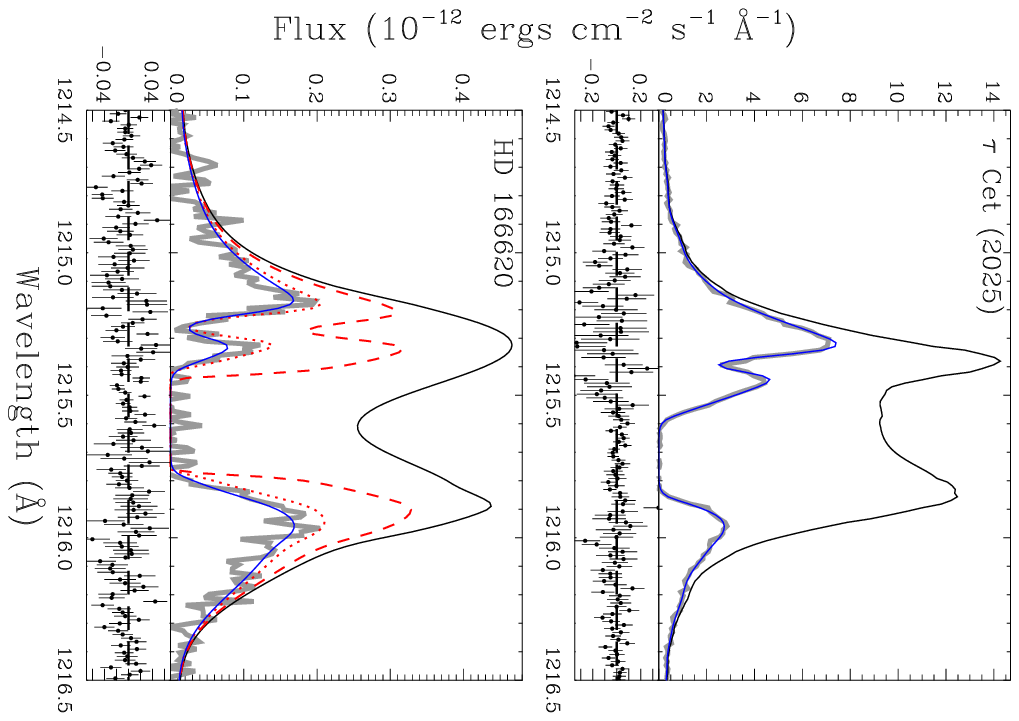}{3.5in}{90}{60}{60}{220}{-50}
\caption{Fits to the H~I Ly$\alpha$ lines of $\tau$~Ceti (for
  the 2025 spectrum) and HD~166620, with residuals shown underneath.
  Solid black lines are the inferred intrinsic stellar line
  profile, and blue lines indicate the ISM absorption line fit to
  the data.  Only a single ISM component is assumed for $\tau$~Ceti,
  while two are assumed for HD~166620 (red lines),
  where the Fe~II fit from Figure~2 is used to constrain
  the two-component fit.}
\end{figure}
     Turning our attention to fitting Ly$\alpha$, the Mg~II
and Fe~II lines assist in the H~I Ly$\alpha$
analyses in a couple ways.  Besides providing information
about the ISM velocity structure, which cannot be discerned in the
much broader H~I and D~I lines, once the ISM absorption
is removed the Mg~II lines provide an excellent first guess for
the intrinsic stellar H~I Ly$\alpha$ profile.  Both Ly$\alpha$
and Mg~II h \& k are very strong chromospheric lines, with similar
profiles in solar spectra.  The self-reversal seen at the center
of the Mg~II lines in Figure~2 is typical of these lines \citep{at24}.
For the narrow ISM Mg~II and Fe~II lines, it is possible to infer a
stellar background by simple interpolation (see Figure~2), but this
is definitely not possible for Ly$\alpha$.  For Ly$\alpha$,
the basic idea is to use the Mg~II k line profile as a first guess
for the shape of the stellar profile.  An initial fit to the ISM
absorption is then performed.  Residuals of that fit are used to
adjust the assumed shape of the background stellar profile, followed
by a new fit to the ISM absorption.  Figure~3 shows our best
Ly$\alpha$ fits after a couple iterations of this.  These are
for the new $\tau$~Ceti spectrum from 2025, and the HD~166620 spectrum.

     Both H~I and D~I lines are fitted simultaneously, taking into
account the two fine-structure constituents of the lines (see Table~3),
and forcing the H~I and D~I lines to have self-consistent velocities
and Doppler parameters.  With H~I and D~I being dominated by
thermal broadening, the latter constraint means
$b({\rm HI})=\sqrt{2}\times b({\rm DI})$.  We also simply assume
a D/H ratio of ${\rm D/H}=1.56\times 10^{-5}$, a value
that has been found to apply throughout the nearby ISM \citep{bew04}.
In this manner, the three fit parameters
of each H~I absorption component are entirely dependent on the
D~I parameters.  Thus, in a single-component fit to the data,
as is the case for $\tau$~Ceti, there are actually only three
free parameters to the entire H~I+D~I absorption fit, and
in Table~3 we therefore only list the H~I parameters.

     For HD~166620, there are two ISM components, and we take this
into account in the Ly$\alpha$ analysis.  Because the two
components are completely blended in H~I and D~I, we use the Fe~II
parameters to constrain the Ly$\alpha$ fit.  We use Fe~II
instead of Mg~II because the Mg~II absorption for HD~166620 is
saturated (see Figure~2), leading to larger uncertainties in the
fit parameters.  In our Ly$\alpha$ fit, we assume the
velocity separation and column density ratio are the same for
H~I+D~I as observed for Fe~II, and we simply assume both
components have the same Doppler parameter.  In this way, even
the two-component HD~166620 Ly$\alpha$ fit has only three
free parameters, and in Table~3 we use parentheses to indicate
that the second ISM component's parameters are not
inferred independently.

     In a number of past analyses, we have sometimes found that
we cannot fit the H~I and D~I absorption in a self-consistent
manner without including an extra H~I absorption component
that is interpreted as being heliospheric and/or astrospheric
in origin, associated with solar/stellar wind interactions
with the ISM \citep{bew05b,bew21}.  We see no evidence for
such extra H~I absorption for $\tau$~Ceti or HD~166620
(see Figure~3), and \citet{ay22} found no evidence
for such absorption for HD~191408 either.

     Figure~1 provides a direct comparison of two separate H~I
Ly$\alpha$ spectra for $\tau$~Ceti.  There is little apparent
difference between the two, though fluxes are slightly lower
for the new 2025 spectrum.  The old spectrum was analyzed by
\citet{bew05b}, and we have here analyzed the new spectrum.
These independent analyses provide an excellent opportunity to
assess the uncertainties in the ISM fit parameters.  The formal
uncertainties in Table~3 are associated with the random errors
induced by the noise in the data, as determined by Monte Carlo
modeling.  However, systematic errors probably dominate,
particularly for the Ly$\alpha$ fits, where uncertainties in
the background stellar profile will be large.  The H~I velocities
of $v=12.34\pm 0.06$ km~s$^{-1}$ and $v=12.59\pm 0.14$ km~s$^{-1}$
are in reasonably good agreement, consistent with the expected
accuracy of the wavelength calibration.  Since we are using the
geocoronal emission as a wavelength calibrator, this accuracy is
defined by how well the centroid of the geocoronal emission can be
measured.  For the noisier 2025 spectrum, the estimated uncertainty
based on the Gaussian fit to the emission is $\pm 0.5$ km~s$^{-1}$.
The Doppler parameters of $b=10.32\pm 0.06$ km~s$^{-1}$ and
$b=10.20\pm 0.19$ km~s$^{-1}$ are in very good agreement.
As for column density, the two measured $\tau$~Ceti values
from Table~3 are $\log N(H)=18.006\pm 0.002$ and
$\log N(H)=18.066\pm 0.005$.  There is a significant 0.06 dex
difference between the two, which is much larger than the random
error estimates, and is presumed to be due to uncertainties in
the reconstructed stellar Ly$\alpha$ profile.
We will return to this issue in Section~5.

     Although we are mostly interested in the chromospheric
Mg~II and Ly$\alpha$ line fluxes, corrected for ISM
absorption, the ISM absorption itself is of some interest, so
we here provide some discussion of the ISM measurements.
For each detected ISM component in Table~3, we have used the
cloud radial velocities and local ISM maps from \citet{sr08}
to identify likely nearby clouds
responsible for the absorption, and these are listed in the
fourth column of Table~3.  Starting with the HD~166620 line
of sight (LOS), the two components are likely identified with the
Mic cloud, with a predicted velocity toward HD~166620 of
$v=-19.82$ km~s$^{-1}$, and the LIC that surrounds the
Sun, with a predicted velocity of $v=-13.35$ km~s$^{-1}$.

     The simpler, single-component $\tau$~Ceti LOS
is potentially intriguing, as there is ambiguity about
whether we are seeing absorption from the LIC, which surrounds
the Sun, or the adjacent G cloud.  Although the Sun is located
within the LIC, it appears to be very near the edge of it,
with the G cloud located on the other side.  The predicted
LIC and G cloud velocities toward $\tau$~Ceti are
$v=11.7$ km~s$^{-1}$ and $v=17.7$ km~s$^{-1}$, respectively
\citep{sr08}.  The Mg~II and Fe~II velocities
are in beteween the two, suggesting there may be a mixture
of the two along the LOS \citep{ahn25}.
This is consistent with the recently expressed view that the
Sun is near or within a compression region that could be
described as being due to a collision between the G and LIC
clouds, possibly associated with a supernova shell
\citep{ps22,cz25,ay25}.

     We here note that the ISM temperature suggested by
the H~I Doppler parameter is intermediate between the G
and LIC clouds, and the depletion values for Mg and Fe are
intermediate as well, consistent with a mixed LOS.
The temperature suggested by the
$b(H)=10.32\pm 0.06$ km~s$^{-1}$ measurement from Table~3
for the older data is $6435\pm 75$~K, assuming purely thermal
broadening.  This is between the $T=7500\pm 1300$~K and
$T=5500\pm 400$~K values for the LIC and G clouds,
respectively \citep{sr08}.  If we assume
the average $\log N(H)=18.04$ value from the two
$\tau$~Ceti measurements, and solar abundances from
\citet{ma09}, the Mg and Fe column densities
from Table~3 suggest ISM dust depletions of
$D(Mg)=-0.52\pm 0.40$ and $D(Fe)=-0.62\pm 0.70$.
The quoted uncertainties are large, but these measurements
are also between the expected LIC and G values:
\citet{sr08} quote LIC and G values of
$D(Mg)=-0.97\pm 0.23$ and $D(Mg)=-0.36\pm 0.35$, respectively,
and $D(Fe)=-1.12\pm 0.10$ and $D(Fe)=-0.54\pm 0.11$,
respectively.  With the G cloud being less dust depleted
than the LIC, a mixture of LIC and G cloud absorption might
also explain why the Mg~II and Fe~II velocities are shifted
more toward the G cloud velocity ($v=17.7$ km~s$^{-1}$)
than H~I (and D~I).

\section{Chromospheric Line Fluxes}

%\begin{deluxetable}{llccccc}
%\tabletypesize{\scriptsize}
%\tablecaption{Absorption Line Fit Parameters\tablenotemark{a}}
%\tablecolumns{7}
%\tablewidth{0pt}
%\tablehead{
%  \colhead{Star}&\colhead{Ion}&\colhead{$\lambda_{rest}$\tablenotemark{b}} &
%    \colhead{ISM} & \colhead{$v$\tablenotemark{c}} & \colhead{$b$} &
%    \colhead{log N} \\ %& \colhead{$\chi^{2}_{\nu}$} \\
%  \colhead{}&\colhead{}&\colhead{(\AA)} & \colhead{Cloud}&\colhead{(km~s$^{-1}$)} &
%    \colhead{(km~s$^{-1}$)} & \colhead{log(cm$^{-2}$)}} % & \colhead{}}
%\startdata
\begin{table}[t]
\begin{center}
%\caption{HST/STIS Targets}
Table 4: Surface Flux Measurements
\small
\begin{tabular}{ccccc} \hline\hline
  Star & \multicolumn{4}{c}{\underline{Surface Fluxes (10$^3$ ergs cm$^{-2}$ s$^{-1}$)}} \\
   & H~I Ly$\alpha$ & Mg~II h+k & C~II $\lambda$1335 & C~IV $\lambda$1548 \\
\hline
HD 166620         &$161\pm 16$ &$522\pm 21$ & $1.06\pm 0.18$ & $1.67\pm 0.29$\\
$\tau$~Cet (old)  &$405\pm 41$ &$743\pm 30$ & $1.84\pm 0.03$ & $1.87\pm 0.04$\\
$\tau$~Cet (2025) &$394\pm 39$ &    ...     & $1.54\pm 0.05$ & $1.46\pm 0.09$\\
HD 191408$^a$     &$259\pm 26$ &$545\pm 22$ & $0.68\pm 0.04$ & $0.77\pm 0.07$\\
Sun (minimum)$^b$ &  $271$     &  $1552$    &      ...       &      ... \\
\hline
\end{tabular}
\end{center}
\small
Notes --- $^a$Flux calibration potentially uncertain due to use of
  the narrow $0.2^{\prime\prime}\times 0.06^{\prime\prime}$ aperture.
  $^b$Solar minimum values from \citet{jll22}.
%NOTE: Assuming 10% errors for Ly-a.  Instead of +/-3,2,4 I'll use 4% for MgII?
\normalsize
\end{table}
%\enddata
%\tablenotetext{a}{Values in parentheses are fixed relative to other component
%  (see text).}
%\tablenotetext{b}{Rest wavelengths of measured lines, in vacuum.}
%\tablenotetext{c}{Central velocity in a heliocentric rest frame.}
%\end{deluxetable}
     With the Mg~II and H~I Ly$\alpha$ lines corrected
for ISM absorption, we can now measure chromospheric line
fluxes from them.  Figure~4 compares the H~I and Mg~II line
profiles of $\tau$~Ceti, HD~166620, and HD~191408; with
integrated line fluxes reported in Table~4.  The fluxes are
displayed and listed in surface flux units assuming the stellar
distances and radii listed in Table~2.  The reported Mg~II
fluxes are for the sum of the h and k lines.
\begin{figure}[t]
\plotfiddle{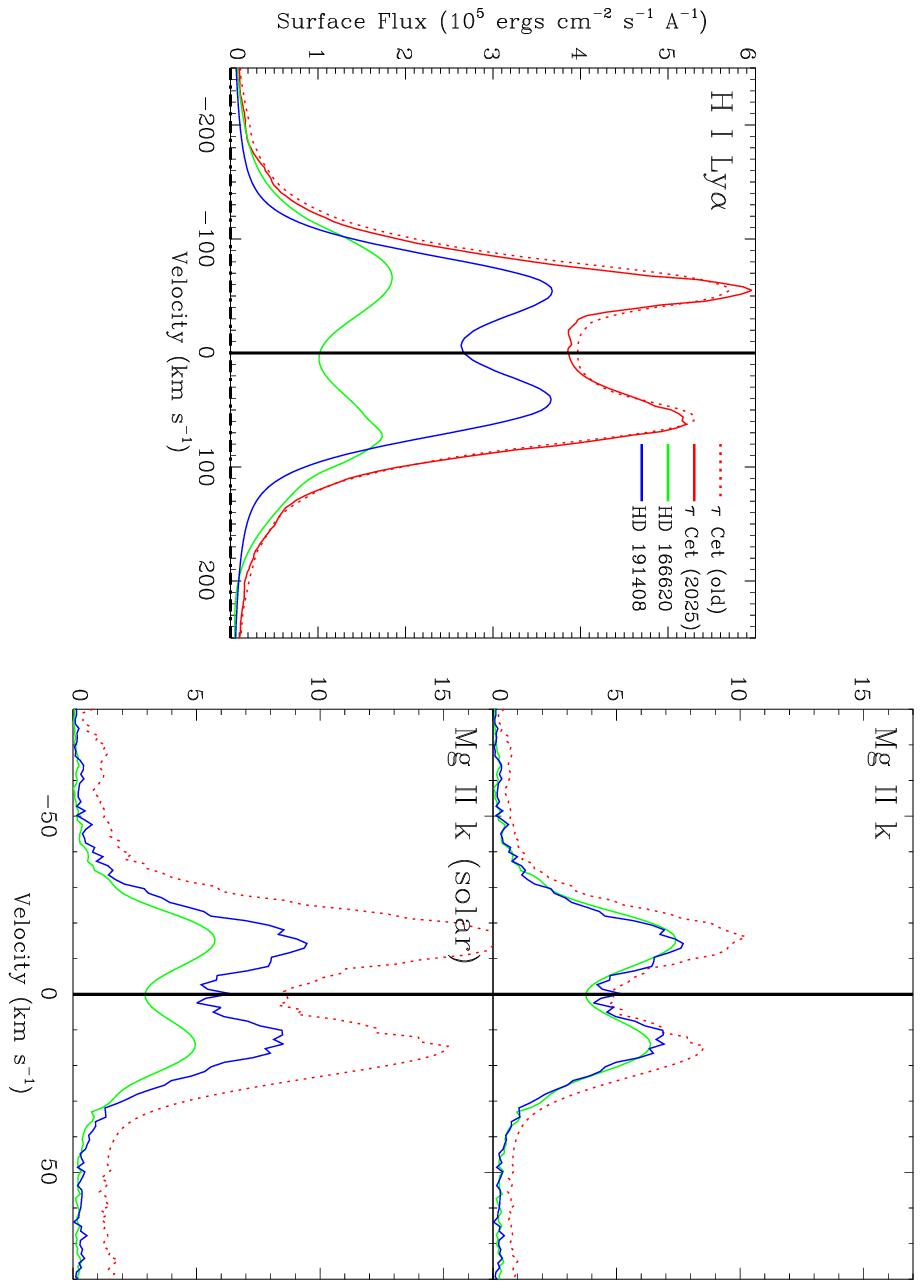}{3.0in}{90}{55}{55}{205}{-65}
\caption{The H~I Ly$\alpha$ and Mg~II k lines of our three G8-K2
  dwarf stars, plotted in surface flux units versus a velocity
  scale in the stellar rest frame.  The bottom panel of the Mg~II
  figure shows the spectra after normalization to a
  solar Mg abundance.}

\end{figure}
     Estimating uncertainties for Ly$\alpha$ fluxes is difficult,
since the uncertainty is dominated by the large correction for
ISM absorption (see Figure~3).  The new $\tau$~Ceti data provide an
opportunity to assess this.  Our analysis of the 2025 $\tau$~Ceti
spectrum has yielded an inferred stellar Ly$\alpha$ profile and line
flux in very good agreement with the previous analysis of the older
E140M spectrum (see Figure~4).  This is both encouraging for the
accuracy of the stellar profile reconstruction, and also is
illustrative of just how steady $\tau$~Ceti's UV line fluxes are,
consistent with the star being a prototypical flat activity star
\citep{acb22}.  Nevertheless, even for nearby stars with
modest ISM absorption flux corrections we expect Ly$\alpha$
flux uncertainties of at least 10\%, so the uncertainties reported
in Table~4 simply assume this uncertainty level.  For Mg~II, flux
uncertainties from noise in the spectra are extremely low, so
in Table~4 we instead quote 4\% uncertainties based on estimates
of flux calibration uncertainties for STIS/E230H spectra
\citep{sh24}.

     A central goal of this study is to compare the chromospheric
line profiles and fluxes for very old, inactive G8-K2~V stars.  The
inferred Ly$\alpha$ line profiles of our three stars are similar,
with similar inferred self reversals near line center, but we do see
significant differences in Ly$\alpha$ line flux in Figure~4.
Our MGM star, HD~166620, appears to have a particularly low flux.
The only G8-K3~V stars that we could find that might have
Ly$\alpha$ surface fluxes this low measured from HST spectra
are Kepler~444 \citep{vb17,jll20} and
Ross~825 \citep{acs19}.  These are also clearly very
old $\sim 10$~Gyr stars, but these are lower spectral resolution
observations toward more distant targets, with correspondingly
larger uncertainties.  Thus, our quoted Ly$\alpha$ surface
flux of $F_{Ly\alpha}=(1.61\pm0.16)\times 10^5$ ergs~cm$^{-2}$~s$^{-1}$
is a plausible estimate of the true minimum value for G8-K2
dwarf stars.  This can be compared with the solar minimum value of
$F_{Ly\alpha}=2.71\times 10^5$ ergs~cm$^{-2}$~s$^{-1}$ from
\citet{jll22}, which is roughly what we see for HD~191408.

     In contrast to H~I Ly$\alpha$, the Mg~II fluxes seen from
our three target stars are nearly identical (see Figure~4).  However,
this consistency may be misleading, as these stars have different
metallicities, with $\tau$~Ceti and HD~191408 being metal poor but
HD~166620 being closer to solar metallicity.  The bottom right
panel of Figure~4 shows the Mg~II lines corrected to solar
abundances based on the [Mg/H] abundances listed in Table~2.
With this correction, we recover a flux progression similar to
that seen for Ly$\alpha$, with fluxes increasing from HD~166620
to HD~191408 to $\tau$~Ceti.  This exercise serves to illustrate one
major advantage of Ly$\alpha$ as a chromospheric diagnostic,
which is that this emission is from the dominant atomic constituent,
hydrogen, and interpretation of Ly$\alpha$ fluxes is therefore
not complicated by abundance issues.

\begin{figure}[t]
\plotfiddle{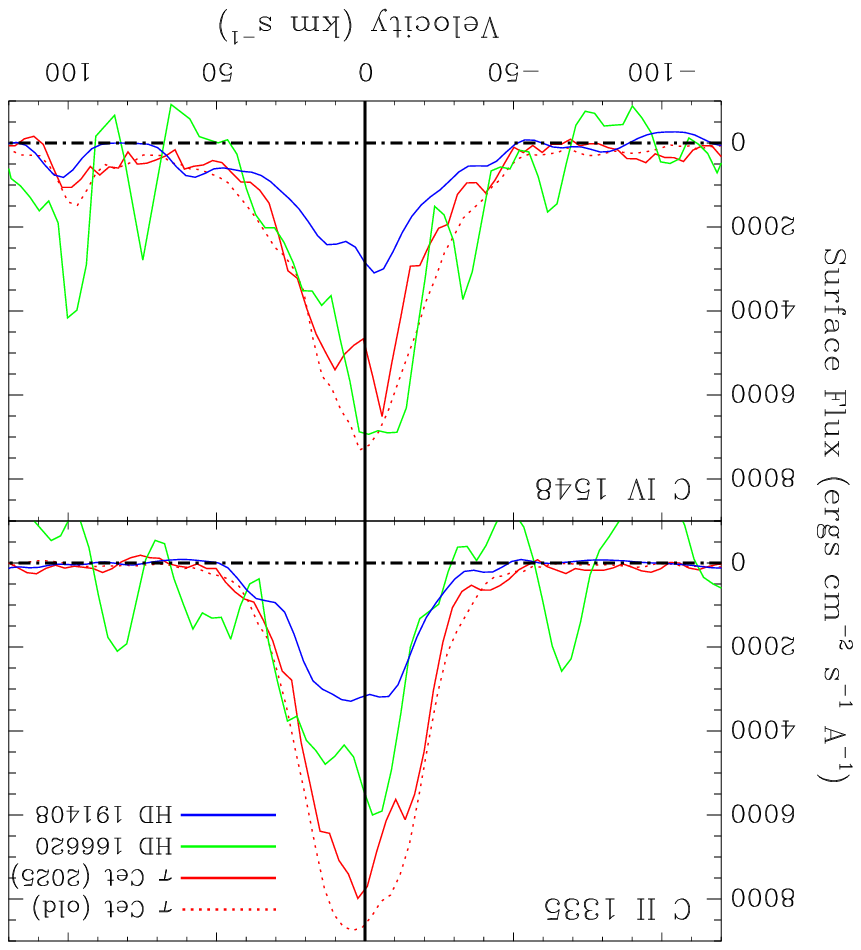}{3.0in}{180}{55}{55}{220}{280}
\caption{Slightly smoothed spectra of the C~II 1335.692~\AA\ and
  C~IV 1548.204~\AA\ lines of our three G8-K2 dwarf stars, plotted
  in surface flux units versus a velocity scale in the stellar
  rest frame.}
\end{figure}
     The H~I Ly$\alpha$ and Mg~II h \& k lines are by far the
strongest chromospheric diagnostics available in our UV spectra,
emanating from plasma with temperatures of $\log T=4.0-4.3$.  In
order to measure activity at higher temperatures, in the
so-called transition region (roughly $4.5<\log T=5.5$), we
also measure line fluxes for the C~II line at 1335.692~\AA,
and the C~IV line at 1548.204~\AA.
The C~II and C~IV lines are much weaker than Ly$\alpha$
and Mg~II, and for HD~166620 and HD~191408 are observed at very
low signal-to-noise (S/N).  The weaker member of the C~IV doublet at
1550.781~\AA\ is in fact undetected, although this is largely
because it is in a different echelle order than the 1548.204~\AA\
line, where the detector background is much higher.  The C~II line
at 1334.519~\AA\ is clearly seen for all our stars, but it is
often affected by ISM absorption, so we focus only on the
C~II 1335.692~\AA\ line.

     The E140M spectrum of HD~191408 is a coaddition of two
exposures from two successive HST orbits, and comparison of the
two exposures reveals a potential problem with data
reduction of low S/N HST echelle spectra.  The spectrum extracted
from the first exposure (dataset odn902010) shows a clear C~IV line,
but this is not the case for the second exposure (dataset odn902030).
Inspection of the echelle image of odn902030 clearly shows C~IV
emission that the standard data reduction is failing to extract.
We note that the detector background is significantly
higher for odn902030 than for odn902010, indicating that this high
background may be responsible for the spectral extraction problems.

     Further study indicates that the standard CALSTIS ``x1d'' spectral
processing, which involves cross-correlation to optimally locate the
individual echelle orders, can fail when the stellar spectrum is weak
and contains only a few faint emission lines.  This is especially true
when the raw echellegrams are dominated by MAMA glow and point-like
hot pixels and particle hits.  For this reason, a new set of
extractions was performed for the low S/N HD~166620 and HD~191408 E140M
data using a fixed global template of order y-locations, adjusted to
match the prominent Ly$\alpha$ feature in order 122 in each exposure,
and employing the standard order traces (``1dt'' file), slit height
(7 pixels), and flanking background extraction widths (5 pixels).
The default scattered-light correction was disabled, as it
provides an unfiltered background trace that can corrupt the net
spectrum if there are strong particle spikes, and instead a
median-filtered and boxcar-smoothed background was applied to avoid
introducing spurious background artifacts into the spectrum itself.
The new extractions show good agreement between the O~I,
C~II, and C~IV features of the separate exposures of HD~191408,
whereas the archive-processed x1d spectra completely misses
C~IV 1548 in the second observation, despite being clearly visible
in the raw image.  The strong Ly$\alpha$ line is not affected at
all by this revised processing.

     The C~II and C~IV line profiles are shown in Figure~5, and the
fluxes are listed in Table~4, where in this case the quoted
uncertainties are simply those associated with the noise in the
spectra.  For $\tau$~Ceti, we find C~II and C~IV 
fluxes that are 15\%--20\% lower in 2025 than they were in
the older 2000 spectrum, representing a significantly larger change
than observed for Ly$\alpha$.  Thus, even the flat-activity
$\tau$~Ceti exhibits some degree of UV variability, particularly
for higher temperature lines.  Note that \citet{pgj04}
and \citet{tra22} provide more detailed studies
of the high-quality $\tau$~Ceti transition region line profiles.

     Surprisingly, for C~II and C~IV the star with the lowest
fluxes seems to be HD~191408, instead of HD~166620 (see Figure~5).
We could normalize the fluxes to solar C abundances, based on the
[C/H] abundances listed in Table~2, analogous to what is done
in Figure~4 for Mg~II.  This mitigates the effect somewhat,
but we still find HD~191408 C~II and C~IV fluxes that are
particularly low.  There is some reason for concern about the flux
calibration of the HD~191408 E140M spectrum, because it was taken
through the narrow $0.2^{\prime\prime}\times 0.06^{\prime\prime}$
aperture instead of the more standard and photometric
$0.2^{\prime\prime}\times 0.2^{\prime\prime}$ aperture.  Nevertheless,
the H~I Ly$\alpha$ fluxes shown in Figures~1 and 4 seem
reasonable for HD~191408, and are identical for the two
aforementioned separate E140M exposures of this star.

     Interestingly enough, it is $\tau$~Ceti that seems to be the
faintest of the three stars in coronal X-rays, if measurements
from ROSAT's High Resolution Imager (HRI) are the guide.
The two ROSAT/HRI exposures of $\tau$~Ceti imply $\log L_X=26.37$
and $\log L_X=26.54$ (in ergs~s$^{-1}$), while for HD~166620 and
HD~191408 ROSAT/HRI measures $\log L_X=26.96$ and
$\log L_X=26.82$, respectively \citep{js04}.
The HD~166620 HRI observation is from October 1996, when the
star was last at an activity cycle maximum \citep{jkl22}.
In any case, for our three stars it is curious that it is
$\tau$~Ceti that seems the least active coronally, HD~191408 that
is least active in the transition region, and HD~166620 that is
least active in the chromosphere.

     Does the particularly low Ly$\alpha$ flux of HD~166620
in Figure~4 imply that stars in MGM phases have chromospheric
fluxes that are very low even compared to stars at activity cycle
minima?  This question is of particular interest because there
has in the past been speculation about what effects solar MGM
phases might have had on Earth's climate.  The Maunder Minimum has
been hypothesized to have contributed to the ``Little Ice Age''
period on Earth, though the much longer duration of the
Little Ice Age casts doubt on this connection \citep{mjo17}.
A lack of direct measurement of solar radiative output during
the Maunder Minimum prevents any definitive determination of
whether any solar radiative changes might have affected Earth's
atmosphere significantly.

     There is potential for UV variability to have an outsized
effect on climate due to its effects on the chemistry of Earth's
upper atmosphere.  Attempts to extrapolate radiative
correlations observed in modern times to the Maunder Minimum
suggest that H~I Ly$\alpha$ fluxes might have been of
order 5\%--10\% lower than solar minimum values
\citep{nak10,cjw18}.  In contrast, our
HD~166620 Ly$\alpha$ flux is about 40\% lower than a typical
solar minimum value, suggesting lower $F_{Ly\alpha}$ is at least
possible.

     A true test of whether an MGM phase significantly
reduces $F_{Ly\alpha}$ would require an HD~166620 Ly$\alpha$
measurement during a typical activity cycle minimum phase for
this star, to compare with the MGM value that we measure here.
Unfortunately, such a measurement is not available, meaning that
it is entirely possible that HD~166620 simply has a particularly
low $F_{Ly\alpha}$ whenever it is in a low activity state.  The Ca~II
H \& K monitoring of the star would seem to support this, as
the Ca~II S-index during its current flat-activity phase
is {\em not} lower than observed during minima of
its former cycling period \citep{acb22,jkl22}.
We have looked at the {\em International Ultraviolet Explorer} (IUE)
archive to see what Mg~II measurements IUE could provide, but
the only relevant observation is a single LWR-HI spectrum from
1983~July~7, which provides a surface flux measurement of
$F_{MgII}=6.3\times 10^5$ ergs~cm$^{-2}$~s$^{-1}$,
about 20\% higher than the STIS value in Table~4.  In 1983,
HD~166620 was nearing a magnetic minimum but was not quite there
\citep{jkl22}, so a slightly higher Mg~II flux in 1983
is consistent with expectations.

\section{Implications of a Weak $\tau$~Ceti Wind}

\subsection{Astrospheric Implications}

\begin{figure}[t]
\plotfiddle{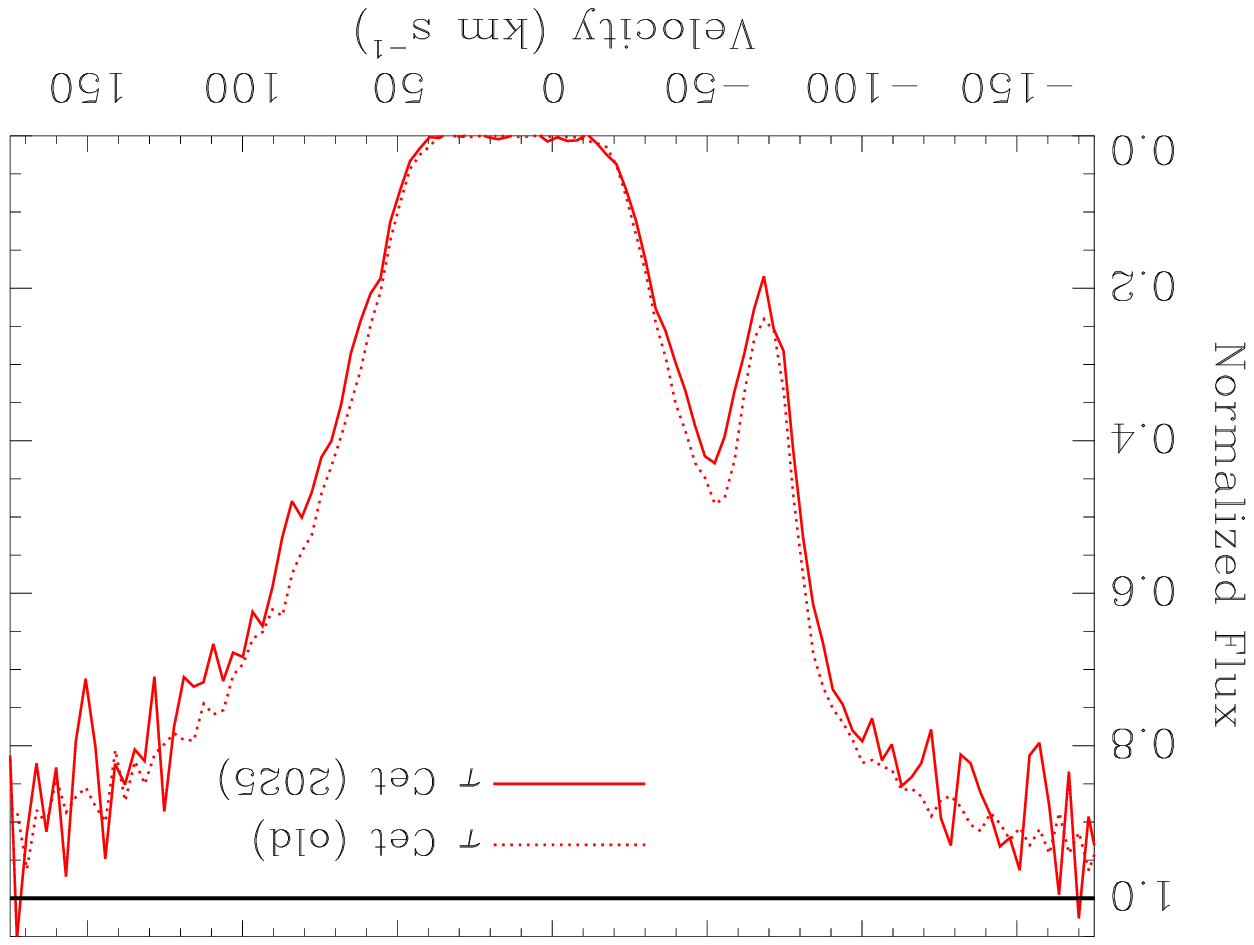}{2.2in}{180}{45}{45}{160}{230}
\caption{Comparison of ISM absorption profiles inferred for the
  H~I+D~I Ly$\alpha$ line of $\tau$~Ceti based on independent
  analyses of both the original STIS/E140M spectrum from 2000,
  and the new STIS/E140M spectrum from 2025.  The spectrum is
  plotted on a heliocentric rest frame.}
\end{figure}
     One aspect of our study is a comparison of two Ly$\alpha$
spectra of $\tau$~Ceti, the older one having been first analyzed
by \citet{bew05b}, and the new one independently analyzed
here (see Figure~3).  In both cases we are able to fit the data
with only ISM absorption, with no evidence for significant
absorption contributions from the stellar astrosphere or our
own heliosphere, as is sometimes seen toward nearby stars
\citep{bew05b,bew21}.  Though undetected, there should in
principle be some astrospheric absorption present on the blue
side of the H~I ISM absorption profile.  In comparing our two
Ly$\alpha$ spectra taken 25 years apart, we look for very
subtle changes in the Ly$\alpha$ absorption
profile that might be indicative of variation in this weak,
undetected astrospheric absorption.  In Figure~6, we directly compare
the inferred H~I+D~I absorption profiles for the two $\tau$~Ceti
spectra.  Our analysis suggests slightly stronger absorption for
the 2025 data than for the old spectrum, with a correspondingly
higher inferred $\log N(H)$ value (see Table~3).

     Is it possible that the higher $\log N(H)$ value for the
new 2025 spectrum could actually be due to a slight increase in
this very weak astrospheric absorption?  This is doubtful, as
such an increase should also result in a slight blueshift of the
H~I absorption, since the blue side of the H~I absorption is where
the astrospheric absorption would reside \citep{bew05b,bew21}.
We see no evidence of such a blueshift, so we interpret the
absorption difference in Figure~6, and the
associated 0.06~dex discrepancy in $\log N(H)$, as being
indicative of systematic errors induced by uncertainties in the
stellar Ly$\alpha$ line profile, which the random
uncertainties quoted in Table~3 fail to quantify.

     Because $\tau$~Ceti is so nearby ($d=3.65$~pc), it is almost
certainly surrounded by partially neutral ISM analogous to that
which surrounds the Sun.  This kind of ISM clearly dominates
within about 7~pc of the Sun, as opposed to the completely ionized
ISM that predominates within the Local Bubble as a whole
\citep{bew18,bew21}.  Therefore, the most likely reason
for the nondetection of astrospheric absorption for $\tau$~Ceti is
simply that its stellar wind is too weak to create a detectable
astrosphere.  The nondetection of astrospheric absorption can
therefore be used to infer a meaningful upper limit for the stellar
mass loss rate.  This has already been done, with an inferred upper
limit of $\dot{M}<0.1$~$\dot{M}_{\odot}$, where the solar mass loss
rate is roughly $\dot{M}_{\odot}=2\times 10^{-14}$ M$_{\odot}$ yr$^{-1}$
\citep{bew18}.  This is in fact the weakest wind yet inferred from any
astrospheric absorption analysis \citep{bew21}.  The
particularly low upper limit implies that the $\tau$~Ceti astrosphere
should in principle have been quite detectable, if the stellar wind
was anywhere near as strong as the solar wind.  The expected
detectability is in part due to the relatively low ISM H~I column
density (and correspondingly narrow ISM absorption).  But also very
important is the high ISM speed of $V_{ISM}=56$ km~s$^{-1}$ seen by
the star, which increases the heating and deceleration of the
ISM H~I as it piles up against the astropause, which should in turn
increase the amount of astrospheric absorption seen blueward of
the ISM absorption.

     The original astrospheric analysis for $\tau$~Ceti was actually
done using hydrodynamic models previously computed for a different
star, 61~Vir, with a similar $V_{ISM}=51$ km~s$^{-1}$ value
\citep{bew18}.  There is value in redoing this
analysis with brand new models computed specifically for $\tau$~Ceti,
with $V_{ISM}=56$ km~s$^{-1}$, in order to see if any revision to the
$\dot{M}$ upper limit is required.  The analysis begins with
hydrodynamic modeling of the astrosphere, for models assuming
different stellar mass loss rates.  As in many past studies
\citep[e.g.,][]{bew21}, the code we use is a 2.5-D axisymmetric,
multi-fluid code that treats the plasma as a single fluid,
but the neutral hydrogen as multiple fluids corresponding to
distinct particle populations created by charge exchange in
different parts of the astrosphere \citep{gpz96}.  It is
crucial to note that the astrospheric models are extrapolated
from a heliospheric model computed from this same code that has
been proven to successfully reproduce heliospheric Ly$\alpha$
absorption \citep{bew00}.  A $\tau$~Ceti astrospheric model
with a solar mass loss rate would simply be the heliospheric model
recomputed for a $V_{ISM}$ changed from the solar value of
$V_{ISM}=26$ km~s$^{-1}$ to the $\tau$~Ceti value
of $V_{ISM}=56$ km~s$^{-1}$ \citep{bew18}.  In order to vary the
assumed stellar mass loss rate, we simply vary the stellar wind
density at the inner boundary.

\begin{figure}[t]
\plotfiddle{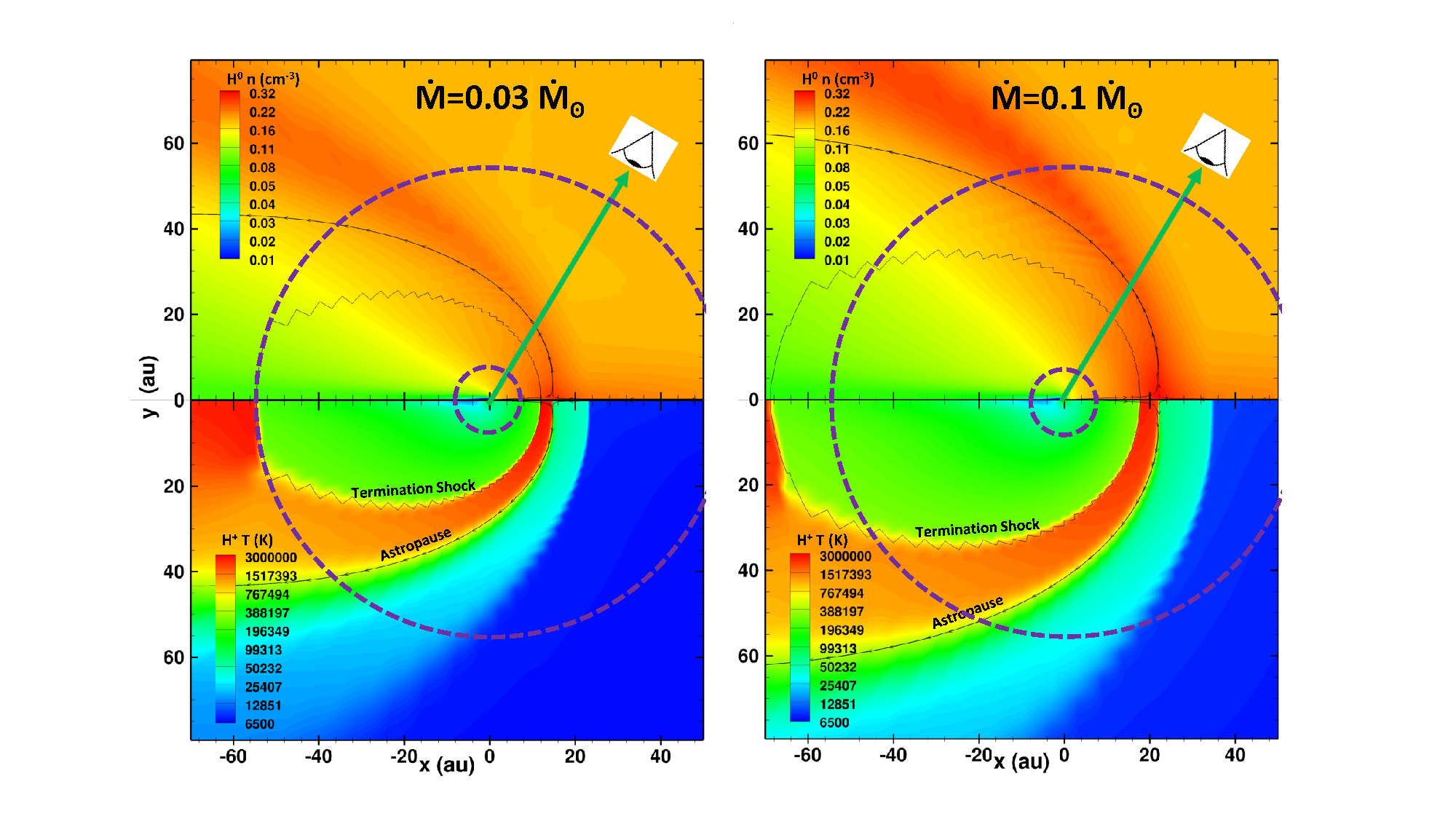}{3.3in}{0}{55}{55}{-255}{-20}
\caption{Hydrodynamic models of the $\tau$~Ceti astrosphere, assuming
  stellar mass loss rates of $\dot{M}=0.03$~$\dot{M}_{\odot}$ (left) and
  $\dot{M}=0.1$~$\dot{M}_{\odot}$ (right).  The top panels show
  neutral H density and the bottom panels show proton temperature.
  The termination shock and astropause boundaries are labeled in
  the plasma temperature panel.  The LOS to the star is
  indicated, at $59^{\circ}$ from the upwind direction of the ISM flow.
  Dashed circles schematically indicate the spatial extent of
  $\tau$~Ceti's debris disk, demonstrating that the disk will be
  at least partly exposed to the ISM outside the astropause,
  regardless of how the disk is actually oriented relative to
  the astrospheric structure.}
\end{figure}
     In Figure~7, we show two models of the $\tau$~Ceti astrosphere,
assuming mass loss rates of $\dot{M}=0.03$~$\dot{M}_{\odot}$ and
$\dot{M}=0.1$~$\dot{M}_{\odot}$.  The
LOS to the star is $59^{\circ}$ from the upwind
direction of the ISM flow in the rest frame of the star.
There are three important boundaries defined by the stellar-wind/ISM
interaction:  the termination shock (TS), where the stellar wind
is shocked to subsonic speeds; the bow shock where the ISM
flow is shocked to subsonic speeds; and between these the
astropause (AP), which is the contact surface separating the
plasma flows of the fully ionized stellar wind and the partially
ionized ISM flow.  The locations of the TS and AP boundaries
are explicitly identified in Figure~7.  From models like those
in Figure~7, we can compute predicted H~I Ly$\alpha$
absorption for our LOS to the star, for comparison
with the data.

\begin{figure}[t]
\plotfiddle{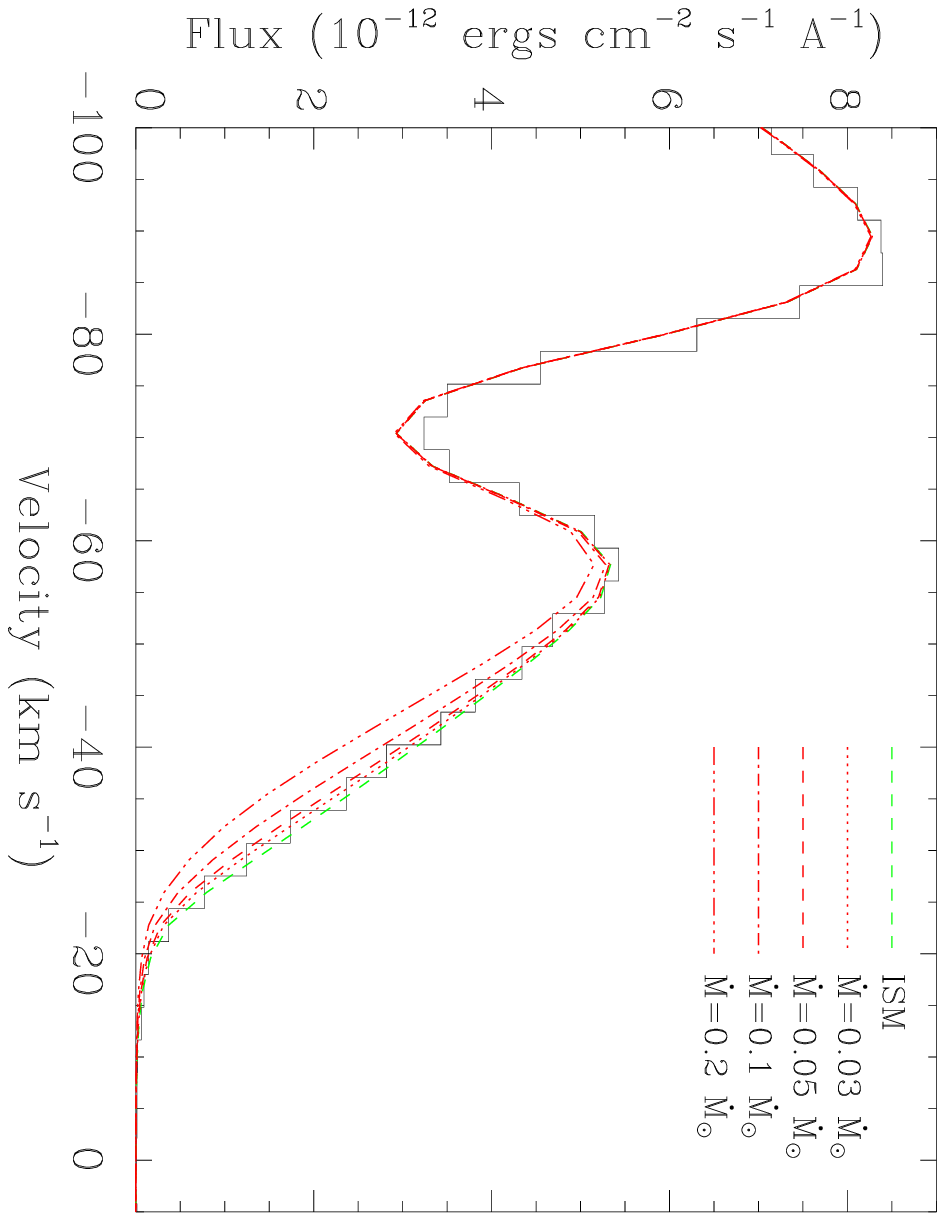}{2.2in}{90}{45}{45}{165}{-45}
\caption{A close-up of the blue side of the H~I Ly$\alpha$
  absorption observed toward $\tau$~Ceti, with the narrow
  absorption at $-73$ km~s$^{-1}$ being ISM D~I.  Predicted
  astrospheric absorption is shown for four models assuming
  stellar mass loss rates of $\dot{M}=0.03-0.20$~$\dot{M}_{\odot}$,
  leading us to quote an upper limit of
  $\dot{M}<0.1$~$\dot{M}_{\odot}$ for $\tau$~Ceti.}
\end{figure}
     In Figure~8, this predicted absorption is compared with the
observed absorption for the old $\tau$~Ceti spectrum from 2000,
which is still used here because it has a much longer exposure time
and therefore significantly higher S/N (see Table~1).
Predicted absorption is shown for 4 models with
mass loss rates ranging from $\dot{M}=0.03-0.20$~$\dot{M}_{\odot}$.
Because there is no actual detected astrospheric absorption,
the goal is to infer an upper limit in $\dot{M}$ based on an
assessment of which models predict absorption strong enough that
it should have been detected.  This assessment is unfortunately
rather subjective.  We believe the $\dot{M}=0.2$~$\dot{M}_{\odot}$
and $\dot{M}=0.1$~$\dot{M}_{\odot}$ models yield absorption that
should have been detectable, but we are not convinced that the
$\dot{M}=0.05$~$\dot{M}_{\odot}$ and $\dot{M}=0.03$~$\dot{M}_{\odot}$
models yield detectable absorption.  Thus, we ultimately conclude
that we still infer an upper limit of $\dot{M}<0.1$~$\dot{M}_{\odot}$,
unchanged from the \citet{bew18} conclusion.

\subsection{Debris Disk Implications}

     Figure~7 indicates just how compact the $\tau$~Ceti astrosphere
must be based on the $\dot{M}<0.1$~$\dot{M}_{\odot}$ measurement.
The TS and AP distances in the upwind direction must be less than
18~au and 22~au, respectively.  These can be compared with the
distances measured in our own heliosphere by the {\em Voyager}
spacecraft, where {\em Voyager~1} encountered the termination shock
and heliopause at distances of 94~au \citep{ecs05} and 121~au
\citep{dag13}, respectively, and {\em Voyager~2} encountered
the termination shock and heliopause at distances of 84~au
\citep{ecs08} and 119~au \citep{lfb19}, respectively.
The compactness of $\tau$~Ceti's astrosphere is due in part to its
weak stellar wind, but is also due to the relatively high ISM
wind speed in the stellar rest frame, $V_{ISM}=56$ km~s$^{-1}$.

     One consequence of $\tau$~Ceti's compact astrosphere
is that $\tau$~Ceti's well-studied debris disk should be at
least partly exposed to the ISM.  Debris disks were first
discovered around nearby stars as IR excesses in
{\em Infrared Astronomical Satellite} (IRAS) data.  Initially,
the detected debris disk stars were dominated by A stars like
Vega and $\beta$~Pic, and a few relatively young FGK stars like
$\epsilon$~Eri.  An exception was $\tau$~Ceti, which was the
first truly old, inactive FGK star to have a detected disk
\citep{hha85}.  Modern IR surveys have actually found the debris
disk phenomenon to be relatively common among FGK stars.
From {\em Herschel} observations, \citet{bm16}
report a detection percentage of 23\% for FGK stars within 15~pc,
with no clear difference between young, active stars and old,
inactive stars.

     Debris disks are believed to be dusty remnants of
protoplanetary disks, with the dust population continuously
renewed by collisions involving planetesimals of various sizes.
The Kuiper Belt is the closest analog of this in
our own solar system, but its total dust mass is well below that
of the debris disk stars.  A recent estimate of the solar system's
dust mass based on {\em New Horizons} data
finds $M_{dust}=8.2\times 10^{-7}$ M$_{\oplus}$ \citep{arp19}.
This is over two orders of magnitude less massive than
$\tau$~Ceti's debris disk, with
$M_{dust}=(2.1\pm 0.7)\times 10^{-4}$ M$_{\oplus}$, and it should
be noted that $\tau$~Ceti's disk is the {\em least} massive in the
debris disk survey of \citet{wsh17}.  This raises an obvious
question about why the debris disk masses of old FGK stars vary
so much.  The $\tau$~Ceti example suggests that age and
metallicity are not major factors, as $\tau$~Ceti has a
disk mass much heftier than that of the Sun despite being
both metal poor and much older (see Table~2).

     One factor that could in principle affect the dust
population of debris disks is exposure to impacts from ISM dust
particles.  Sandblasting of larger bodies in debris disks by
ISM dust could contribute to dust renewal in debris disks, or
at least alter the characteristics of the dust in the disk.
The possibility of ISM sandblasting affecting debris disk dust
was considered in the early days of debris disk research.
However, it was largely dismissed because for A stars,
which account for a large fraction of detected debris disks,
radiation pressure prevents the lighter ISM dust particles from
coming anywhere near the star \citep{pa97},
and for coronal FGK stars like the Sun, the astrospheric
coccoons created by coronal winds serve the same purpose
\citep{im06}.  Only the largest ISM dust particles can
penetrate our Sun's heliopause, and such particles are too few
to be important.  Thus, since the Kuiper Belt is comfortably encased
inside the heliopause, ISM dust impacts should have little
effect on the dust population in our own solar system.

     However, these arguments against the relevance of ISM
exposure do not apply to $\tau$~Ceti's debris disk.
Radiation pressure will be unimportant for a G8~V star, and
we find here that $\tau$~Ceti's astrosphere is so compact that
it will not protect the disk from ISM exposure either, at
least not fully.  Observations from {\em Herschel} and the
Atacama Large Millimeter/submillimeter Array (ALMA) suggest a
debris disk located roughly $6-55$~au from the
star \citep{sml14,mam16}.  This debris disk
extent is shown schematically in Figure~7.  Although no
attempt is made to properly orient the disk relative to
the astrospheric structure and our LOS to the star in this
figure, it is clear that the outermost part of the disk
will be at least partly beyond the astropause regardless of
orientation, and therefore exposed to the ISM.  It is worth
pointing out that the disk is observed roughly face-on from
Earth, with a tilt angle of about $i=35^{\circ}\pm 10^{\circ}$
from the plane-of-sky \citep{sml14}.  A full 3-D orientation
of the disk relative to the astrosphere and LOS could in
principle be inferred from this, if combined with
the estimated disk position angle of
$PA_{disk}=105^{\circ}\pm 10^{\circ}$ \citep{sml14}, and the
position angle of the ISM flow vector in the stellar rest frame,
which we infer from the stellar motion and local ISM flow vector
to be about $PA_{ISM}=96^{\circ}$.  This is, however,
complicated by ambiguity about which direction the disk is
tilted relative to the plane of sky.

     It is worthwhile to consider whether there are other debris
disk stars with Ly$\alpha$ observations that indicate astrospheres
compact enough for the disk to be exposed to the ISM.  An
astrospheric detection for a star not only indicates the size of
the astrosphere, but also proves that the star is surrounded by
ISM that is at least partly neutral, presumably containing a
population of dust available to impact debris disks through
sandblasting effects.  This distinction is important, as much of
the Local Bubble that surrounds the Sun is actually fully ionized
ISM that probably lacks a significant dust constituent.
We are aware of only three stars with detected astrospheric
absorption that also have reported debris disks:  $\epsilon$~Eri,
$\xi$~Boo, and 61~Vir \citep{bm16,wsh17,bew21}.
The astrospheres of $\epsilon$~Eri
and $\xi$~Boo are significantly larger than our heliosphere, and
are therefore likely to be large enough to fully contain their debris
disks \citep{bew02,bew05a}.  Thus, those disks should not
experience significant exposure to the ISM.  This is not the case
for the G5~V star 61~Vir, however, which has a weak wind of only
$\dot{M}=0.3$~$\dot{M}_{\odot}$, and which sees a high ISM wind
speed of $V_{ISM}=51$ km~s$^{-1}$ \citep{bew05a}.

\begin{figure}[t]
\plotfiddle{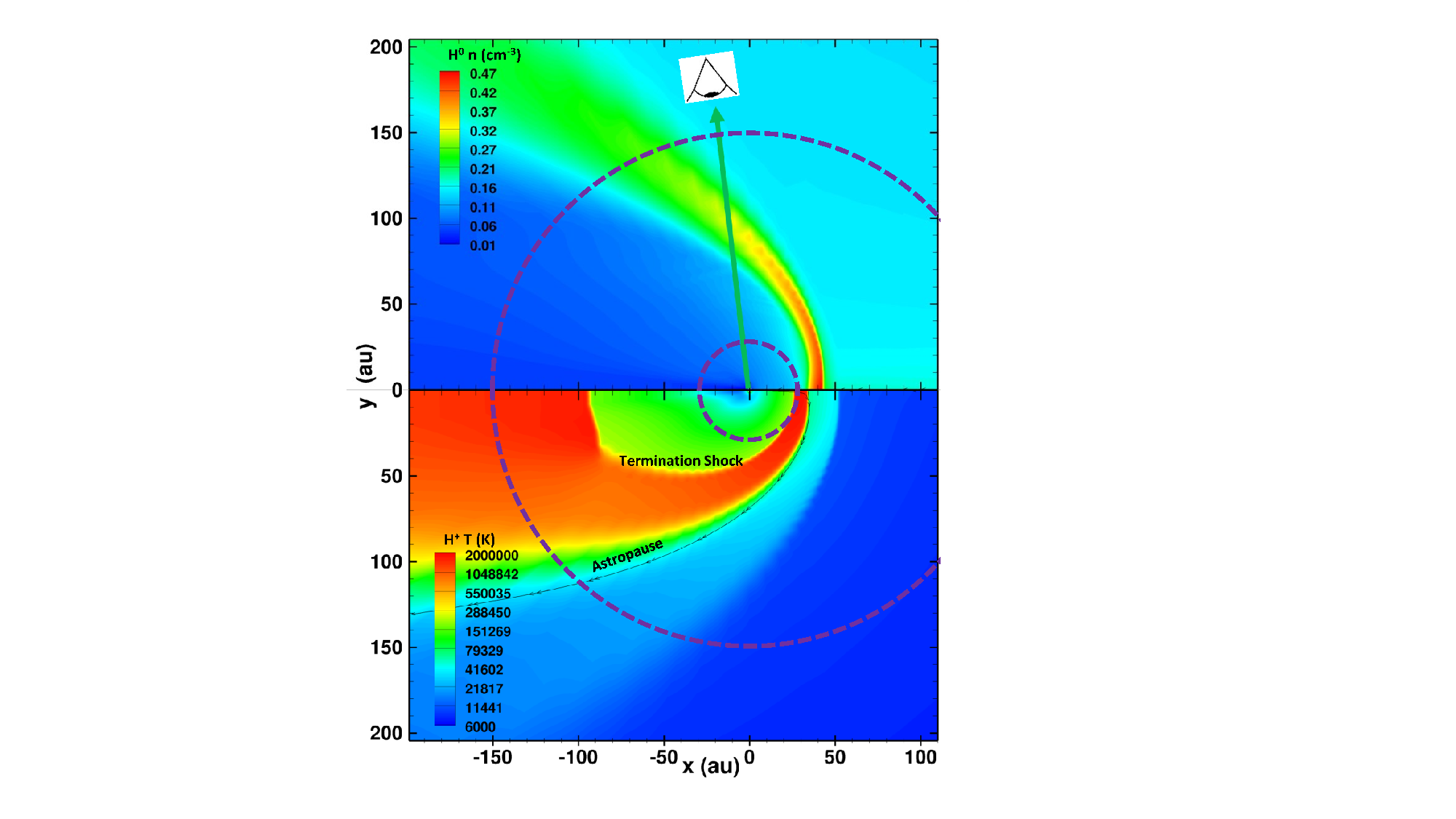}{3.3in}{0}{55}{55}{-245}{-20}
\caption{The best-fit $\dot{M}=0.3$~$\dot{M}_{\odot}$ hydrodynamic model
  for the astrosphere of 61~Vir from Wood et al.\ (2005a) is shown,
  with neutral H density in the top panel and plasma temperature in
  the bottom.  As in Figure~7, dashed circles schematically
  indicate the extent of 61~Vir's debris disk, demonstrating that the
  disk will be at least partly exposed to the ISM.}
\end{figure}
     In Figure~9, we show a figure analogous to Figure~7,
but for 61~Vir instead of $\tau$~Ceti.  The best-fit
astrospheric model from \citet{bew05a} is replicated, and the
$\sim 30-153$~au debris disk extent illustrated schematically
\citep{mcw12,sm17}.  As is the case for $\tau$~Ceti,
the outermost part of the debris disk is at least partly
exposed to the ISM outside the astropause.  Unlike $\tau$~Ceti,
we are observing this debris disk nearly edge-on,
with $i=82^{\circ}\pm 4^{\circ}$ and
$PA_{disk}=59^{\circ}\pm 5^{\circ}$ \citep{sm17}.  For completeness,
we note that from 61~Vir's stellar motion we estimate
$PA_{ISM}=28^{\circ}$.

     If sandblasting from ISM dust is an important process, there
should in principle be observable consequences for the debris
disk properties of $\tau$~Ceti and 61~Vir.  The spectral energy
distributions (SEDs) of debris disks are often modeled with
simple single temperature blackbodies, but with a modification
factor of $(\lambda_0/\lambda)^{\beta}$ beyond the turnover
wavelength $\lambda_0$, in order to correct for emission
inefficiencies of small grains at higher wavelengths
\citep{gmk12}.  In SED fits, $\lambda_0$ and $\beta$
can be free parameters of the fit.  If dust properties in a
certain population of debris disks were uniquely affected by
a particular physical process, such as ISM sandblasting, this
in principle might reveal itself in unusual $\lambda_0$
or $\beta$ parameters.

     Inspection of existing surveys does not imply any clear
oddity about $\tau$~Ceti or 61~Vir in this respect
\citep{wsh17,bs18}.
However, uncertainties in $\lambda_0$ and $\beta$ are large,
and the two parameters often degenerate, leading \citet{wsh17}
to simply quote wide ranges of possible
values for most stars.  Furthermore, for FGK
stars there is also a concern that stellar chromospheres are
contributing to excess emission above that of the photosphere
at long wavelengths, complicating this kind of analysis even
further \citep{jv14,mam16}.
Theoretical modeling is needed to identify observable diagnostics
of ISM sandblasting in debris disk observations, in order to
test if the disks of $\tau$~Ceti and 61~Vir are demonstrably
affected by this process.

     Finally, besides $\tau$~Ceti, it is worth noting that there
is another nearby G8~V star, HD~61005 ($d=36.6$~pc), with a
debris disk that has been proposed to be affected by ISM
exposure \citep{dch07,hlm09}.  The
unusually asymmetric appearance of this disk in HST
images led to its moniker ``The Moth''.  Although identical
in spectral type to $\tau$~Ceti, HD~61005 is different in
being young and very active, and its debris disk is more
massive by over two orders of magnitude \citep{wsh17}.
Interaction with the surrounding ISM is one possible
way to explain the asymmetries seen for The Moth, but this
would seem to require much higher ISM densities than observed
around the Sun.  Such high densities are very rare within
the Local Bubble, but we do note that HD~61005 is located in the
ring of high ISM column densities that has been interpreted
as an extended compression region in the local ISM, where
high ISM densities would be more likely \citep{ps22,cz25,ay25}.
The debris disks of HD~32297 (A0~V) and HD~15115 (F4~IV)
provide additional examples of relatively nearby disks with
asymmetries that may be induced by ISM
interaction \citep{sr07,jhd09,tb19}.
However, disk shaping by massive planets with eccentric,
inclined orbits is another possible way to explain disk
asymmetries \citep{tme16}.

\section{Summary}

     We have presented and analyzed new HST UV spectra of
HD~166620 and $\tau$~Ceti.  The former represents the
first UV measurements for a star established to be in an MGM
phase, in which a previous clear stellar activity cycle has
ceased.  We compare the UV properties of HD~166620 with
those of $\tau$~Ceti and HD~191408, two stars similar in
both spectral type and old age.
Our findings can be summarized as follows:
\begin{enumerate}
\item Analysis of the stellar chromospheric H~I Ly$\alpha$
  and Mg~II h \& k lines requires correcting for ISM absorption,
  so a necessary analysis of the ISM absorption along the
  HD~166620 and $\tau$~Ceti lines of sight is provided.
\item We find that our MGM star, HD~166620, has H~I Ly$\alpha$
  surface fluxes significantly lower than those of $\tau$~Ceti
  and HD~191408, and nearly a factor of two lower than observed
  for the Sun at solar minimum.  The Mg~II h \& k lines show
  a similar pattern only if Mg~II fluxes are normalized from
  stellar to solar photospheric abundances.
\item The very low Ly$\alpha$ and Mg~II fluxes for HD~166620
  could in principle imply particularly low chromospheric
  activity for stars in MGM phases, but this conclusion is
  dubious since the Ca~II S-index used to monitor HD~166620's
  activity for decades does not show S-index values in
  HD~166620's MGM phase that are lower than observed during
  stellar minima observed when the star was cycling \citep{acb22,jkl22}.
  Unfortunately, no UV spectra of HD~166620 are available to
  see if that also applies to UV fluxes at such minima.
\item The comparative chromospheric activity levels of our
  three old, inactive G8-K2 stars are not necessarily echoed in
  the transition regions or coronae of these stars.  We find
  that it is HD~191408 that actually has the lowest transition
  region C~IV $\lambda$1548 flux of the three stars.  And it
  is $\tau$~Ceti that is faintest in coronal X-ray emission,
  based on archival ROSAT/HRI data.
\item For $\tau$~Ceti, our new Ly$\alpha$ data are compared
  with an older observation from 2000.  We find very little
  difference in stellar Ly$\alpha$ flux, consistent with
  $\tau$~Ceti being a flat-activity star with no known cyclic
  variation in activity.  Our analysis of the new spectrum
  measures an ISM H~I column density 0.06 higher than the
  value previously measured from the old spectrum, which is
  presumed to indicate the magnitude of uncertainty induced
  by uncertainties in the reconstructed stellar emission
  line profile.
\item Astrospheric Ly$\alpha$ absorption is undetected for
  $\tau$~Ceti, and using new models we confirm a previous
  upper limit for the star's mass loss rate of
  $\dot{M}<0.1$~$\dot{M}_{\odot}$, the weakest wind yet
  inferred from the Ly$\alpha$ diagnostic.
\item The $\tau$~Ceti astrosphere is compact enough that
  the star's debris disk must be at least partly exposed
  to the ISM, with potential ramifications for the dust
  population of the disk due to sandblasting by ISM dust
  grains.  We note that this is also true for one other
  star, 61~Vir, which has both a known debris disk
  and detected astrosphere.
\end{enumerate}

\acknowledgments

We would like to thank Dr.\ T.\ Ayres for useful discussions about
STIS data processing and UV spectra of old main sequence stars.
Support for HST program GO-17793 was provided by NASA through an award from
the Space Telescope Science Institute, which is operated by the Association
of Universities for Research in Astronomy, Inc., under NASA contract
NAS 5-26555.  BEW also acknowledges financial support from
the Office of Naval Research.  TSM acknowledges support from NASA
grant 80NSSC25K7563 and NSF grant AST-2507890.  This research has
made use of the SIMBAD database, operated at CDS, Strasbourg, France.
All the HST data used in this paper were obtained from
the Mikulski Archive for Space Telescopes (MAST) at STScI,
with the specific observations available at
\dataset[DOI: 10.17909/wtt9-z260]{https://doi.org/10.17909/wtt9-z260}.

\end{document}